\documentclass[fleqn,usenatbib]{mnras}

\usepackage{newtxtext,newtxmath}

\usepackage[T1]{fontenc}

\DeclareRobustCommand{\VAN}[3]{#2}
\let\VANthebibliography\thebibliography
\def\thebibliography{\DeclareRobustCommand{\VAN}[3]{##3}\VANthebibliography}

\usepackage{graphicx}	
\usepackage{amsmath}	
\usepackage{makecell}
\usepackage[table]{xcolor}
\usepackage{etoolbox}
\usepackage{pgf} 
\definecolor{high}{HTML}{dd0f00}  
\definecolor{low}{HTML}{9bdf01}  
\newcommand*{\opacity}{90}
\newcommand*{\minvallogg}{0.01}
\newcommand*{\maxvallogg}{0.22}
\newcommand*{\minvallogl}{0.00}
\newcommand*{\maxvallogl}{0.21}

\newcommand{\gradientlogg}[1]{
    \ifdimcomp{#1pt}{>}{\maxvallogg pt}{#1}{
        \ifdimcomp{#1pt}{<}{\minvallogg pt}{#1}{
            \pgfmathparse{int(round(100*(#1/(\maxvallogg-\minvallogg))-(\minvallogg*(100/(\maxvallogg-\minvallogg)))))}
            \xdef\tempa{\pgfmathresult}
            \cellcolor{high!\tempa!low!\opacity} #1
    }}
}
\newcommand*{\minvalteff}{0.00}
\newcommand*{\maxvalteff}{0.30}

\newcommand{\gradientteff}[1]{
    \ifdimcomp{#1pt}{>}{\maxvalteff pt}{#1}{
        \ifdimcomp{#1pt}{<}{\minvalteff pt}{#1}{
            \pgfmathparse{int(round(100*(#1/(\maxvalteff-\minvalteff))-(\minvalteff*(100/(\maxvalteff-\minvalteff)))))}
            \xdef\tempa{\pgfmathresult}
            \cellcolor{high!\tempa!low!\opacity} #1
    }}
}
\newcommand{\gradientlogl}[1]{
    \ifdimcomp{#1pt}{>}{\maxvallogl pt}{#1}{
        \ifdimcomp{#1pt}{<}{\minvallogl pt}{#1}{
            \pgfmathparse{int(round(100*(#1/(\maxvallogl-\minvallogl))-(\minvallogl*(100/(\maxvallogl-\minvallogl)))))}
            \xdef\tempa{\pgfmathresult}
            \cellcolor{high!\tempa!low!\opacity} #1
    }}
}

\newcommand{\ApplyGradient}[1]{%
  \iftoggle{inTableHeader}{#1}{
    \ifdim #1 pt > \MidNumber pt
        \pgfmathsetmacro{\PercentColor}{max(min(100.0*(#1 - \MidNumber)/(\MaxNumber-\MidNumber),100.0),0.00)} %
        \hspace{-0.33em}\colorbox{green!\PercentColor!yellow}{#1}
    \else
        \pgfmathsetmacro{\PercentColor}{max(min(100.0*(\MidNumber - #1)/(\MidNumber-\MinNumber),100.0),0.00)} %
        \hspace{-0.33em}\colorbox{red!\PercentColor!yellow}{#1}
    \fi
  }}
  
\title[O-type Stars Stellar Parameter Estimation]{O-type Stars Stellar Parameter Estimation Using Recurrent Neural Networks}

\author[Flores R. M. et al.]{
Miguel Flores R.,$^{1}$\thanks{E-mail: miguel.flores.r@gmail.com (MFR)}
Luis J. Corral,$^{1,2}$
Celia R. Fierro-Santillán$^{3}$
and Silvana G. Navarro$^{1,2}$
\\
$^{1}$\small Centro Universitario de Ciencias Economico Administrativas (CUCEA), Universidad de Guadalajara, 45100 Zapopan, Jal., México.\\
$^{2}$\small Instituto de Astronomía y Meteorología (IAM), Universidad de Guadalajara, 44130 Guadalajara, Jal., México\\
$^{3}$\small Escuela Nacional Colegio de Ciencias y Humanidades, Plantel Sur (ENCCH Sur), Universidad Nacional Autónoma de México (UNAM), 04500, CDMX, México.
}

\date{Accepted XXX. Received YYY; in original form ZZZ}

\pubyear{2022}

\begin{document}
\label{firstpage}
\pagerange{\pageref{firstpage}--\pageref{lastpage}}
\maketitle

\begin{abstract}
In this paper, we present a deep learning system approach to estimating luminosity, effective temperature, and surface gravity of O-type stars using the optical region of the stellar spectra.
In previous work, we compare a set of machine learning and deep learning algorithms in order to establish a reliable way to fit a stellar model using two methods: the classification of the stellar spectra models and the estimation of the physical parameters in a regression-type task. Here we present the process to estimate individual physical parameters from an artificial neural network perspective with the capacity to handle stellar spectra with a low signal-to-noise ratio (S/N), in the $<$20 S/N boundaries. 
The development of three different recurrent neural network systems, the training process using stellar spectra models, the test over nine different observed stellar spectra, and the comparison with estimations in previous works are presented.
Additionally, characterization methods for stellar spectra in order to reduce the dimensionality of the input data for the system and optimize the computational resources are discussed.
\end{abstract}




\section{Introduction}
Processing big astronomical databases is a current challenge that has been approached from different perspectives \citet{Bin_2020, Villavicencio_2020, Sharma_2020, Bu_2020, Sharma_2019, Fierro_2018, Minglei_2017, Li_2017, Dafonte_2016, Sander_2015, Navarro_2012, Teimoorinia_2012,  Snider_2001}; therefore, with the recent growth of the artificial intelligence methods is natural to think in the implementation of an autonomous systems. In this context, Artificial Neural Networks (ANNs) have proven to be a very useful tool \citet{2022_param_est, Villavicencio_2020,  Bu_2020,Sharma_2019, Navarro_2012}. 

The growing development of a diversity of artificial neural network structures has generated a natural interest of implement it in multiple astrophysics projects. The flexibility of the ANNs to be trained with different tensor-like data structures, their capacity to handle large volumes of data without the necessity of storing a big set of measures like different machine learning algorithms, and the resistance to noise and lack in the data, has presented an advantage in the stellar spectra classification, as well as morphological classification, segmentation of stars and galaxies, and also, the estimation of stellar parameters \citet{2022_param_est, Bu_2020, Bin_2020, Fierro_2018,Dafonte_2016, Recio-Blanco_2006}. 

Early projects has shown highly accurate results on the determination of physical parameters from stellar spectra using modern deep learning techniques. In the context of ANNs, these types of algorithms have achieved estimations over spectra with a high level of noise, near 20 S/N, and explore different ways of improving the parameter estimation accuracy below that limit \citet{2022_param_est, Villavicencio_2020, Navarro_2012}.

This high noise limit has been a threshold where other methods indicate a substantial increase of error as is show in \citet{Recio-Blanco_2006}. In \citet{Worley_2012},is mentioned that in some cases a considerable amount of spectra in their database was rejected from the system and physical parameters do not estimated because of the quality in the spectra. And in \citet{ULySS}, the impact of low S/N on the estimation over the analyzed spectra is not addressed.

Therefore, with the increasing astronomical observations in a more automatic way, with modern and space telescopes, generating databases in the big data limit, the automatic computational systems are an essential tool in order to process and analyze spectral data in a fast and reliable way. According to this, the proposal on this work consist of an deep learning algorithm implementation as a part of a system that estimates fundamental stellar parameters such as $T_{eff}(K)$, $log$ $g$, $log(L/L\textsubscript{\(\odot\)})$ in an accurate way. Additionally, in this project, we propose a characterization method to reduce the dimension of the training database, in order to improve the efficiency of the system and reduce the time and computational resources that a big database required.

This paper is organized as follows. Section \ref{section2} describes the synthetic spectra database used in this work. Also, we introduce the method proposed for the dimensionality reduction process, and preprocessing procedures needed.
In Section \ref{section3}, we introduce the artificial neural networks and the structure we develop for this project. Section \ref{section4} describes the estimation process, the performance evaluation results, and the comparison of the estimation with previous works.
Finally, a brief summary of the results is made in Section \ref{section5}.

\section{Database Collection and Processing}
\label{section2}
\begin{figure}
    \centering
    \includegraphics[scale=0.28]{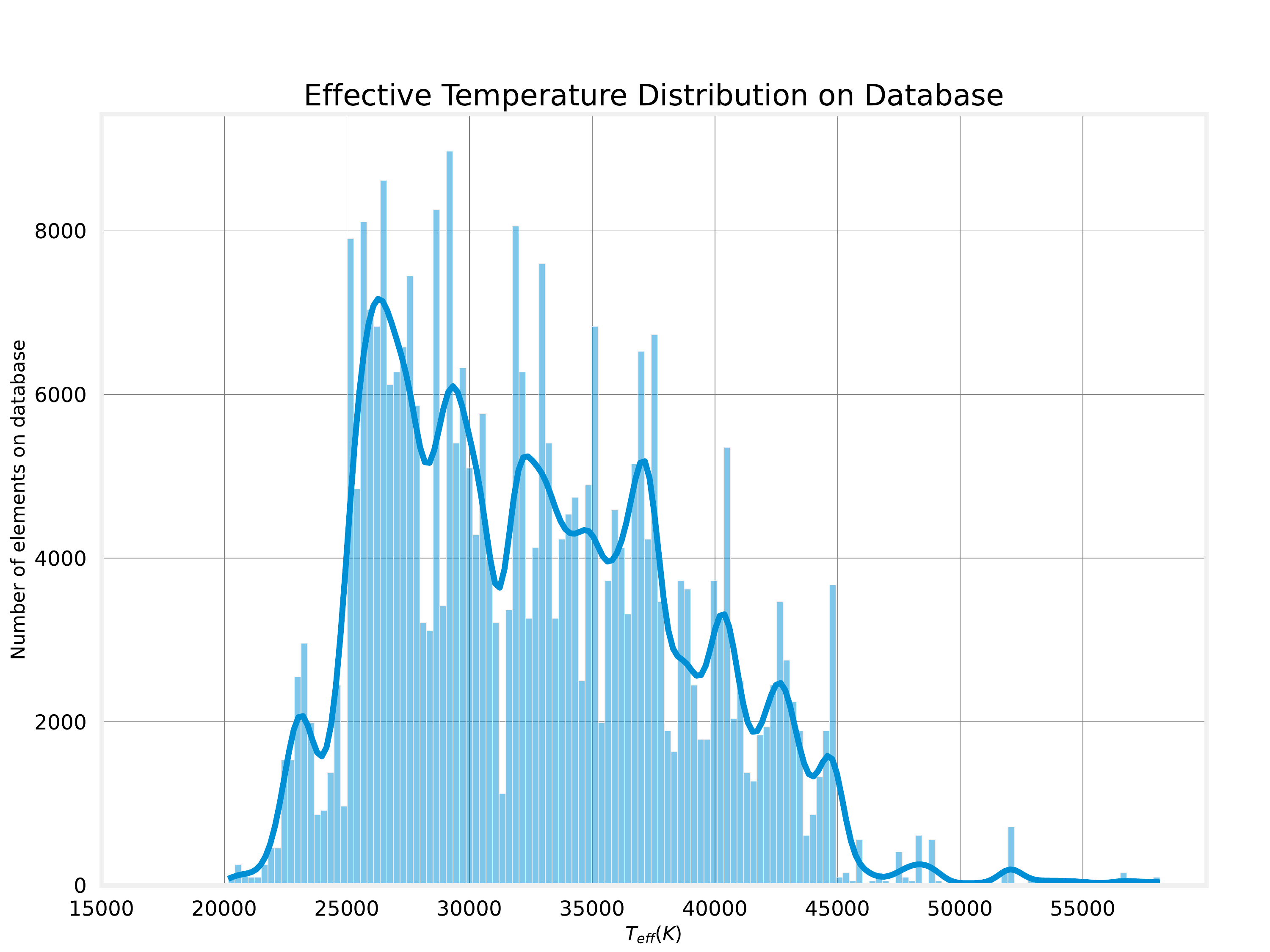}
    \caption{Effective temperature distribution over the 4,517 models in current database.}
    \label{fig: teff_dist}
\end{figure}

\begin{figure}
    \centering
    \includegraphics[scale=0.28]{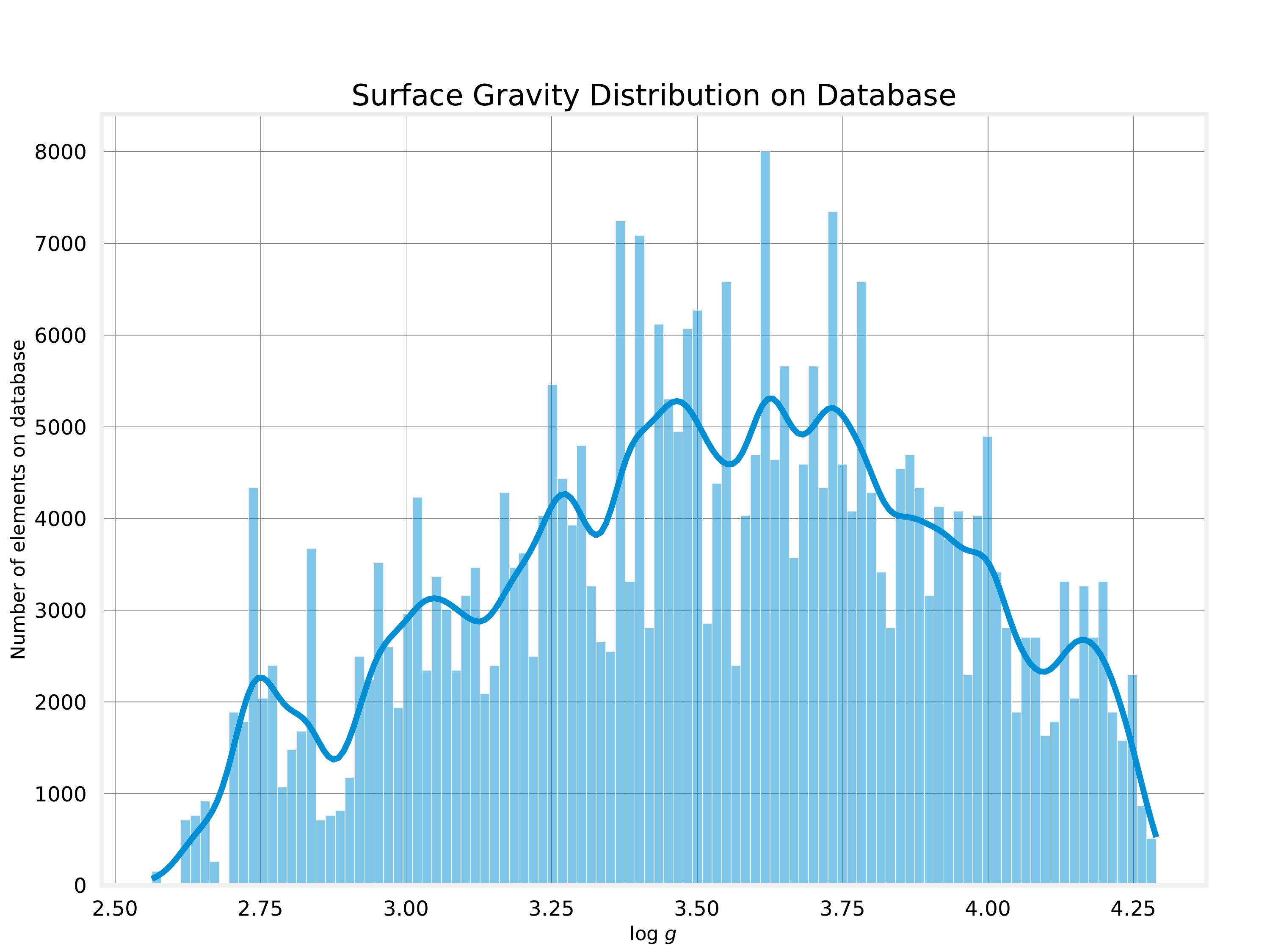}
    \caption{Surface gravity distribution over the 4,517 models in current database.}
    \label{fig: logg_dist}
\end{figure}

\begin{figure}
    \centering
    \includegraphics[scale=0.28]{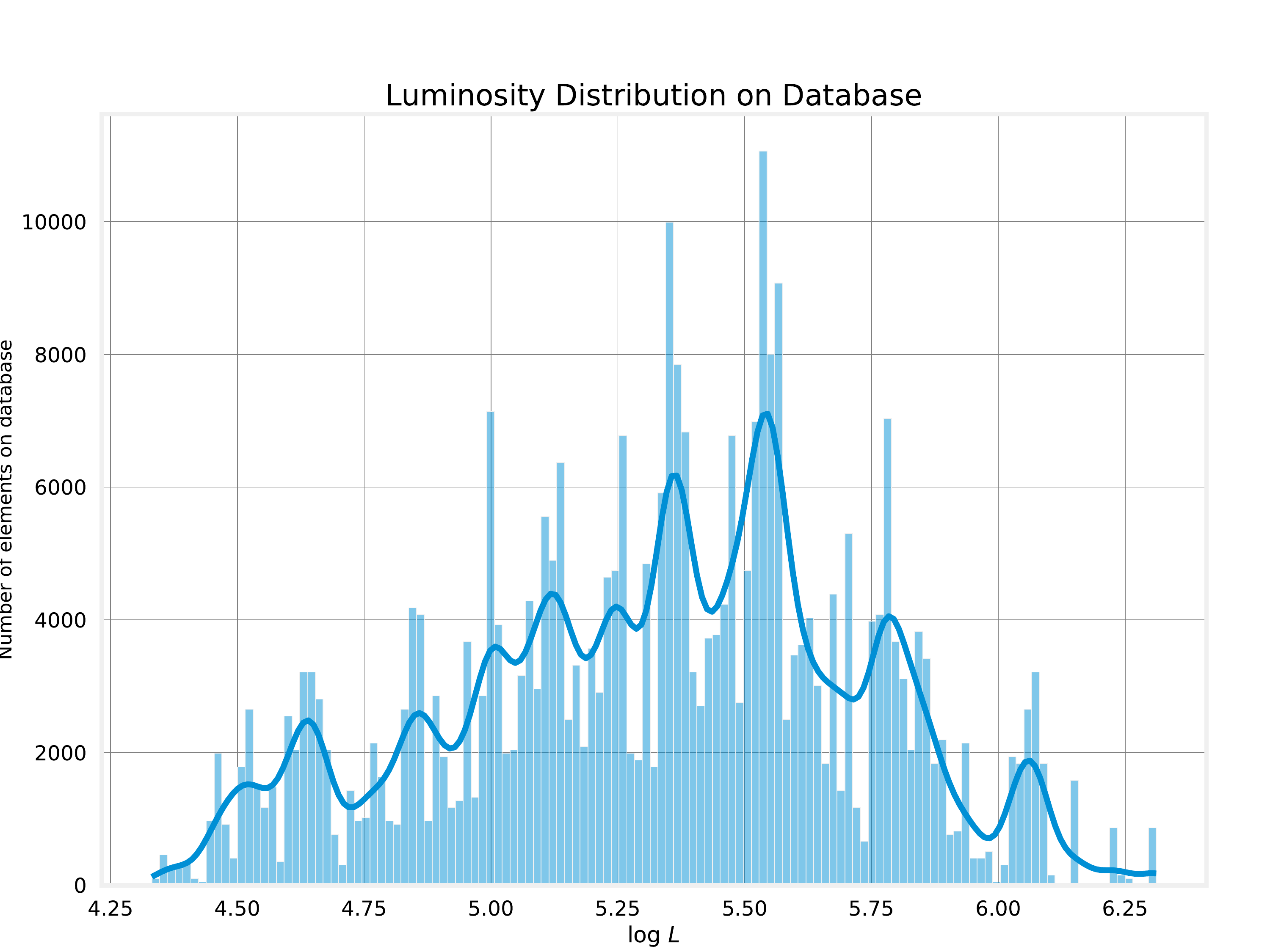}
    \caption{Luminosity distribution over the 4,517 models in current database.}
    \label{fig: logl_dist}
\end{figure}

The starting database used for this work consist on 4517 O-type star models develop by \citet{Zsargo_2020}. The models cover stars with mass from near 9 to 120 $M\textsubscript{\(\odot\)}$, $T_{eff}(K)$ from 20,160 to 58,140, $log(L/L\textsubscript{\(\odot\)})$ from 4.33 to 6.31, and $log$ $g$ from 2.56 to 4.28, with the distribution of Effective, Temperature, Surface Gravity, and Luminosity   as shown on Figures \ref{fig: teff_dist}, \ref{fig: logg_dist}, and \ref{fig: logl_dist}. Each synthetic model is represented by a spectrum in the optical region ($\lambda \lambda$ 3500-7000 \AA) of 100,001 points with uniform step of 0.035 \AA. We modify each spectrum adding white gaussian noise at different levels, in order to obtain spectra with S/N ratio between 150 an 20. This step was done in order to provide noisy spectra to the deep learning algorithm, and generate the capacity to deal with observed spectra which naturally have noise.

Since the results obtained in different projects using artificial neural networks (ANNs) (\citet{2022_param_est, Bu_2020, Sharma_2020, Villavicencio_2020, Sharma_2019, Navarro_2012, Snider_2001}), we proposed a system based on the recurrent network structure, and following the idea of reduce the necessity of high computational resources, we explore the simple network structures and non-network structures needed to complete the tasks.

For the training process, we use a final database with a total of 140,027 synthetic spectra, the 4,517 base models with 30 different levels of signal to noise ratio (S/N) plus the clean synthetic spectrum. 

\subsection{Dimensionality Reduction}
Develop an expert-type system that estimates a set of physical parameters can start with the method that will compare a synthetic spectra generated by a model and an observed spectra. This type of challenge can be reach using either the full spectra, or, by establish a group of variables that characterize the full spectra. The second method can be see as a dimensionality reduction task, which take the advantages of the ANNs to distinguish between a small set of characteristic variables and reduce the computational resources compared to use a signal with a large amount of points.

We proposed a method to reduce the amount of data that gets into the ANNs as the input features. Starting on the system proposed in \citet{Fierro_2018}, where a set of equivalent width (EW) measures is needed, we expand the number of absorption/emission lines measurements following the same process presented in \cite{FloresR}, in order to have information of lines in the red part of the visual spectrum, and also lines of heavier elements such as SiIII, SiIV, etc. The lines of SiII (4128) and SiIII (4552), which are useful to restrict the Temperature of O-type stars, were excluded because most of the spectra are faint, with an equivalent width of thousandths or less. Additionally, the SiII (4128) line has several neighboring lines, which increases the uncertainty of its equivalent width measurement. 

We choose 34 absorption lines that we think can be the minimal number to get high accuracy for both approaches. These 34 lines are shown in Table \ref{tab: Lines}.  For these lines, we check the possible neighbor lines that can be blended.

\begin{table}
\small
\setlength\tabcolsep{1pt}
    \centering
    \begin{tabular}{c|c|c}
    \makecell{Wavelength \\ Centroid (\AA)} & Possible Ions in the Line & Line index\\
    \hline
        3750  & NIV, H12, NIII, OIII (x2) & $L_{1}$  \\
        3770  & H11, SiIV, OIII, NeII, OII & $L_{2}$ \\
        3798  & NeII, OIII, SiII, H10, OII & $L_{3}$ \\
        3820  & HeI  & $L_{4}$\\
        3835  & OII H9  & $L_{5}$\\
        3889  & OII, CII, H8  & $L_{6}$\\
        3926  & HeI, SiIII  & $L_{7}$\\
        3934  & CaII  & $L_{8}$\\
        3970 & HeI, HeII, CaII, H$\epsilon$  & $L_{9}$\\
        4026  & HeI  & $L_{10}$\\
        4058  & NIV & $L_{11}$\\
        4088  & SiIV & $L_{12}$ \\
        4101  & CaII, H$\delta$  & $L_{13}$\\
        4143 & HeI  & $L_{14}$\\
        4200  & HeII  & $L_{15}$\\
        4212 & SiIV  & $L_{16}$\\
        4340 & OII (x2) H$\gamma$ & $L_{17}$ \\
        4388  & HeI  & $L_{18}$\\
        4471  & HeI & $L_{19}$ \\
        4481  & MgII & $L_{20}$\\
        4541  & HeII  & $L_{21}$\\
        4813  & SiIII & $L_{22}$ \\
        4861  & HeII, H$\beta$ & $L_{23}$\\
        4943  & OII (x2) & $L_{24}$\\
        4987  & NII  & $L_{25}$\\
        5005  & NII (x6) & $L_{26}$\\
        5016  & HeI & $L_{27}$\\
        5048  & NII, HeI & $L_{28}$\\ 
        5801  & CIV  & $L_{29}$\\
        5812  & CIV  & $L_{30}$\\
        6381  & NIV & $L_{31}$\\
        6527  & HeII  & $L_{32}$\\
        6563  & HeII, H$\alpha$  & $L_{33}$\\
        6683  & HeII & $L_{34}$\\
    \end{tabular}
    \caption{Total lines needed as an input for the ANNs in the new approach.}
    \label{tab: Lines}
\end{table}

Because the EW does not provide information about the morphology of the line, wich can be very useful to accurate estimate fundamental parameters as the rotational velocity, we create a secondary approach where the input of the networks, are not only the EW, we also add a set of profiles of the measured lines.

\begin{figure*}
    \centering
    \includegraphics[scale=0.3]{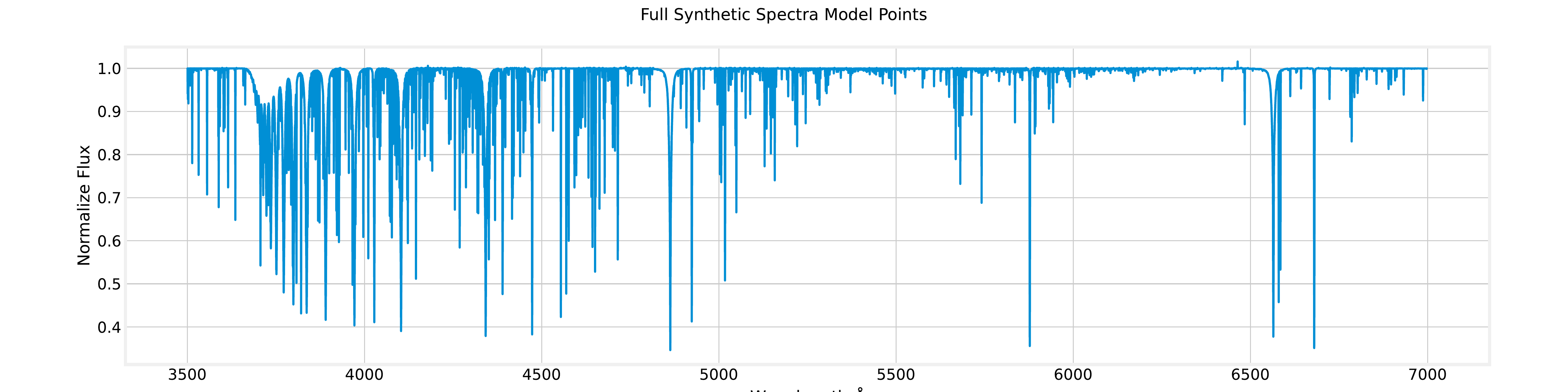}
    \caption{Example of a full synthetic spectra of 100,001 points.}
    \label{fig: FSS}
\end{figure*}

\begin{figure*}
    \centering
    \includegraphics[scale=0.3]{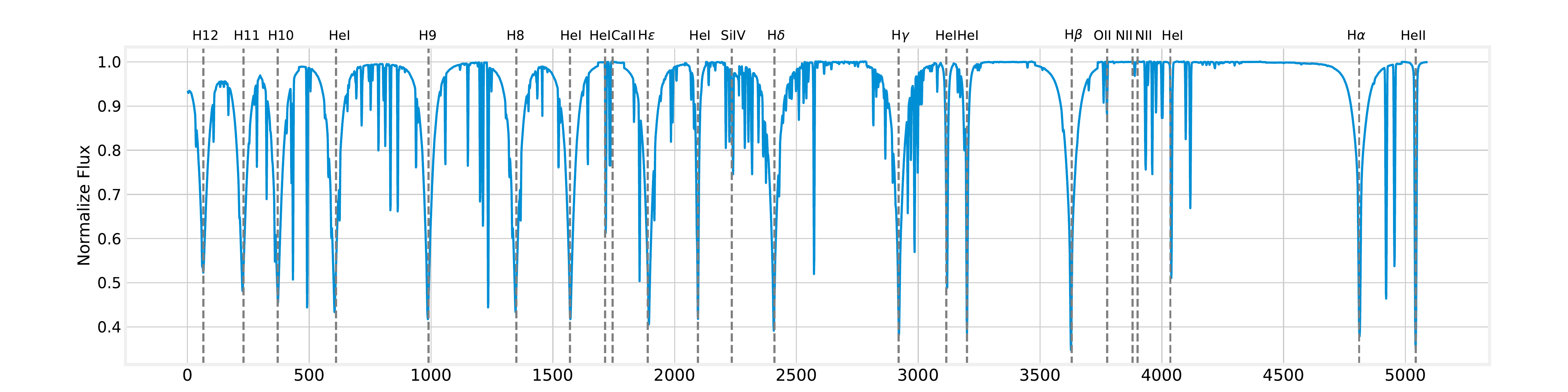}
    \caption{Synthetic spectra of 5092 points (same spectra on Figure \ref{fig: FSS} after the dimensionality reduction process). Some of the lines shown in Table \ref{tab: Lines} are marked for reference.}
    \label{fig: RSS}
\end{figure*}

We define two ways to establish the line profiles needed to get more information about the spectra and do not interfere with one of our statements of avoid the necessity of super-computers.
The first idea was to use all the profiles of the non He or H Ions, having a total of 12 profile lines. The second one, get the top $N$ line profiles with higher correlation with every physical parameter. As a result, we will have an independent set for $T_{eff}(K)$, $log(L/L_o)$, and $log$ $g$. We chose the $N$ as 6 for this work, since was observed as an average break point where the correlation start to decrease in all the physical parameters. The 6 line profiles for every physical parameter are shown in Table \ref{tab: Top6}.
Moreover, the wave length resolution of the observed spectra we use in this work is 0.1226 \AA, this is around 3.5 times less points per spectra (see Section 4). Therefore, in order to have the same tensor lengths for the training database and the observed spectra, we use an interpolation method to reduce the synthetic spectra in a 4 to 1 points that will give us a similar resolution respect to the observed spectra. After the resolution decrease process, we extract all the 34 line profiles we use to measure the EW and then create a signal-type vector without the wavelength relation and from this new group of line profiles we extract the group of profiles for the two methods we mentioned previously in this section. The final process in this dimensionality reduction, it is to match the final length with the observed spectra, for this task we create a simple routine in order to finally have the same number of points for each line profile.
In Figures \ref{fig: FSS}, \ref{fig: RSS}, we show an example of full points synthetic spectrum, and the same spectrum after the process of resolution reduction and the extraction of the line profiles regions for all the 34 lines shown in Table \ref{tab: Lines}, and some lines tagged for a reference on where the lines should be situated. In the case of only using a smaller set of these 34 lines, the chosen lines will be set in a similar way generating and smaller signal-type vector.

{\renewcommand{\arraystretch}{1.2}
\begin{table}
\setlength\tabcolsep{4pt}
    \centering
    \begin{tabular}{c c}
     Physical parameter & Group of lines by Line index in Table \ref{tab: Lines}\\
    \hline
       $T_{eff}(K)$ &  $\{ L_{1}, L_{3}, L_{7}, L_{8}, L_{9}, L_{10} \}$\\
       $log$ $g$   &  $\{L_{21}, L_{28}, L_{30}, L_{31}, L_{33}, L_{34} \}$\\
       $log(L/L\textsubscript{\(\odot\)})$ & $\{L_{26}, L_{28}, L_{29}, L_{31}, L_{32}, L_{33} \}$
        \\
    \end{tabular}
    \caption{Top 6 line profiles with higher correlation per physical parameter over the models data base.}
    \label{tab: Top6}
\end{table}}

\begin{figure}
    \centering
    \includegraphics[scale=0.4]{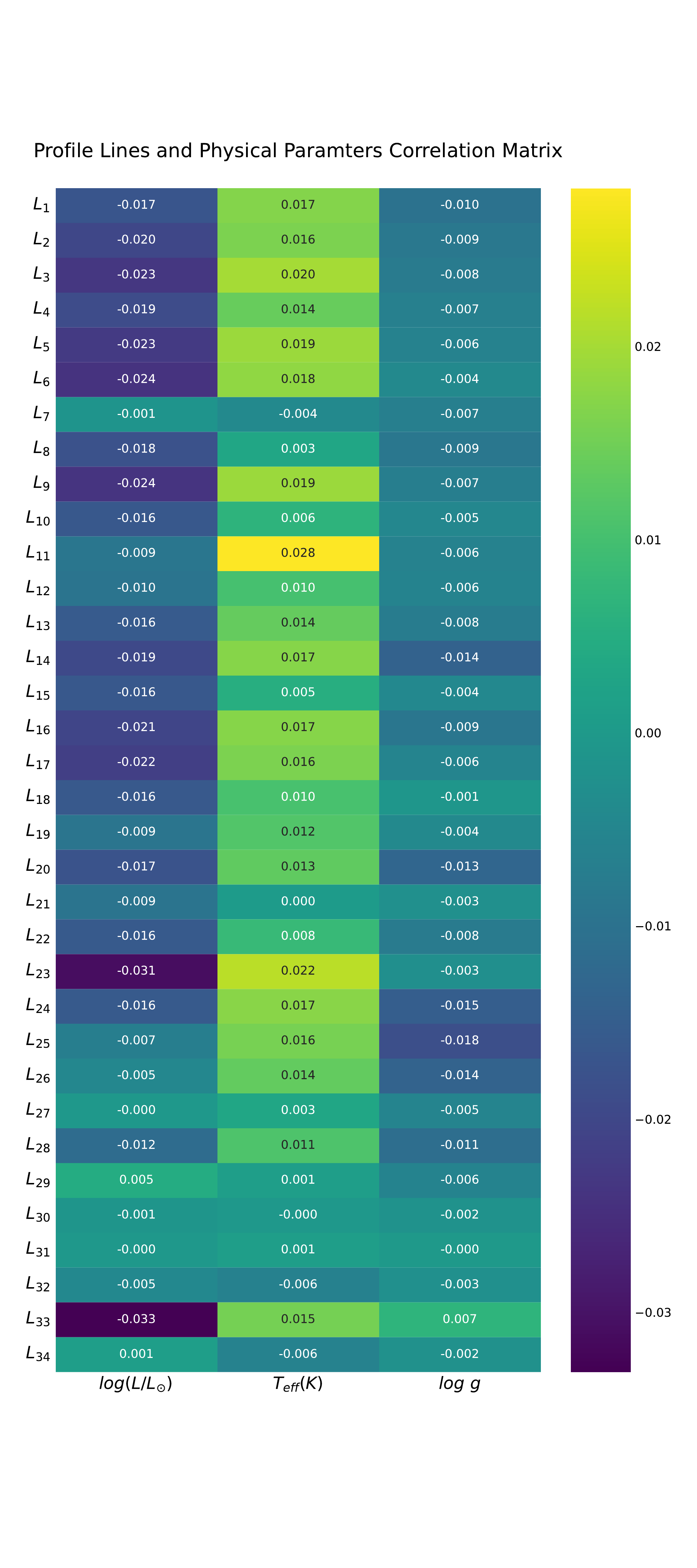}
    \vspace{-2cm}
    \caption{Correlation matrix between the profiles lines of the 34 lines in \ref{tab: Lines} and the Effective Temperature, Surface Gravity, and Luminosity.}
    \label{fig: lines_corr}
\end{figure}

\section{Artificial Neural Networks}
\label{section3}
Because we have two different types of inputs we try to choose a network type capable to deal with the signal-type input from the different profiles and the numerical sequences from the equivalent width measures. 

The election of this type of network structure is based on the type of input data, which is a numerical vector with an specific sequence for each spectrum, and the values will change in relation with the noise in the spectrum. Additional to this, the system needs to have is the capacity to distinguish between the nearby models. One classic solution to the systems with very similar possible outputs is the connection between the neurons and also the feedback connections.

\subsection{Recurrent Neural Networks}
Recurrent neural networks (RNNs) belong to a type of neural networks naturally suited to solve sequential data problems. As a principal property, these class of deep learning algorithms use information from previous inputs in the current stage of input and output processing. While classical or simple deep neural networks commonly assume independence between inputs and outputs, then, for RNNs the output depends on the previous elements in the sequence.
Therefore, it is clear to think of this type of network as one of the best options for our purposes.

\begin{figure}
    \centering
    \includegraphics[scale=0.47]{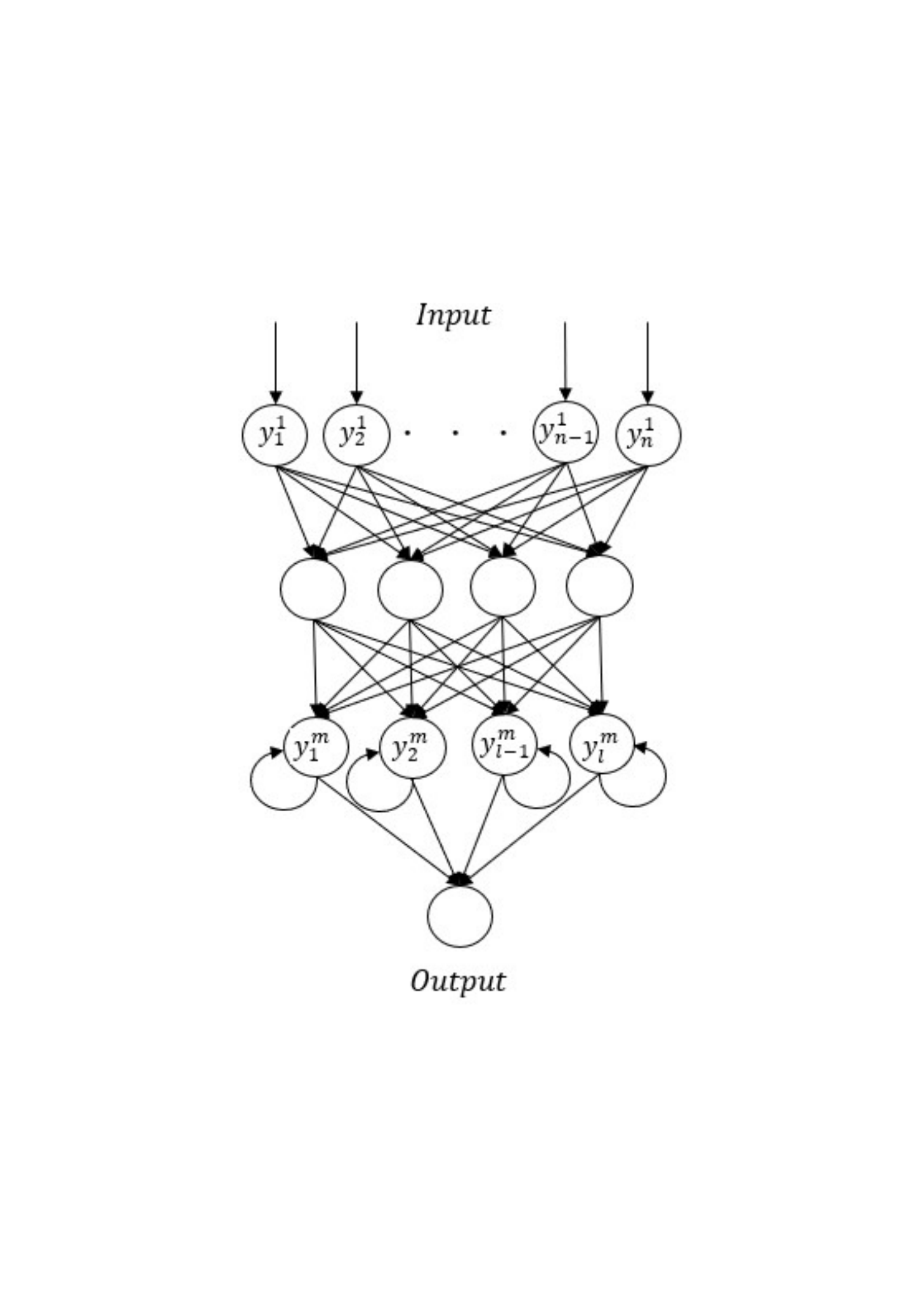}
    \vspace{-2cm}
    \caption{Diagram of a recurrent neural network, which consists of an input layer, a hidden layer and an output layer.}
    \label{fig: rnn_diagram}
\end{figure}

As mentioned before, for this work we established a full system where every physical parameter will be estimated for a individual network, besides, every network will be train using 3 different input tensor. The first one, a rank one tensor of only a sequence of 34 EW, and for the second and third approach a rank three tensor define by the 34 EW for the first entry, and the line profiles for the second entry of 914 point for the non He or H lines, and for the top 6 with higher correlation:  677 points for $log$ $g$, 588 points for $T_{eff}(K)$, and 506 points for $logL$.

Because of the different input tensor dimension, the structure of the RNN will be a functional-type of network, with two paths, that can see as two parallel RNNs. The first path will process the 34 EW measures, and the second path will process a signal-type input created by the Non He/H profile lines or the top $N$ line profiles with higher correlation, as shown on Figure \ref{fig: RSS}.

\begin{figure}
    \centering
    \includegraphics[scale=0.4]{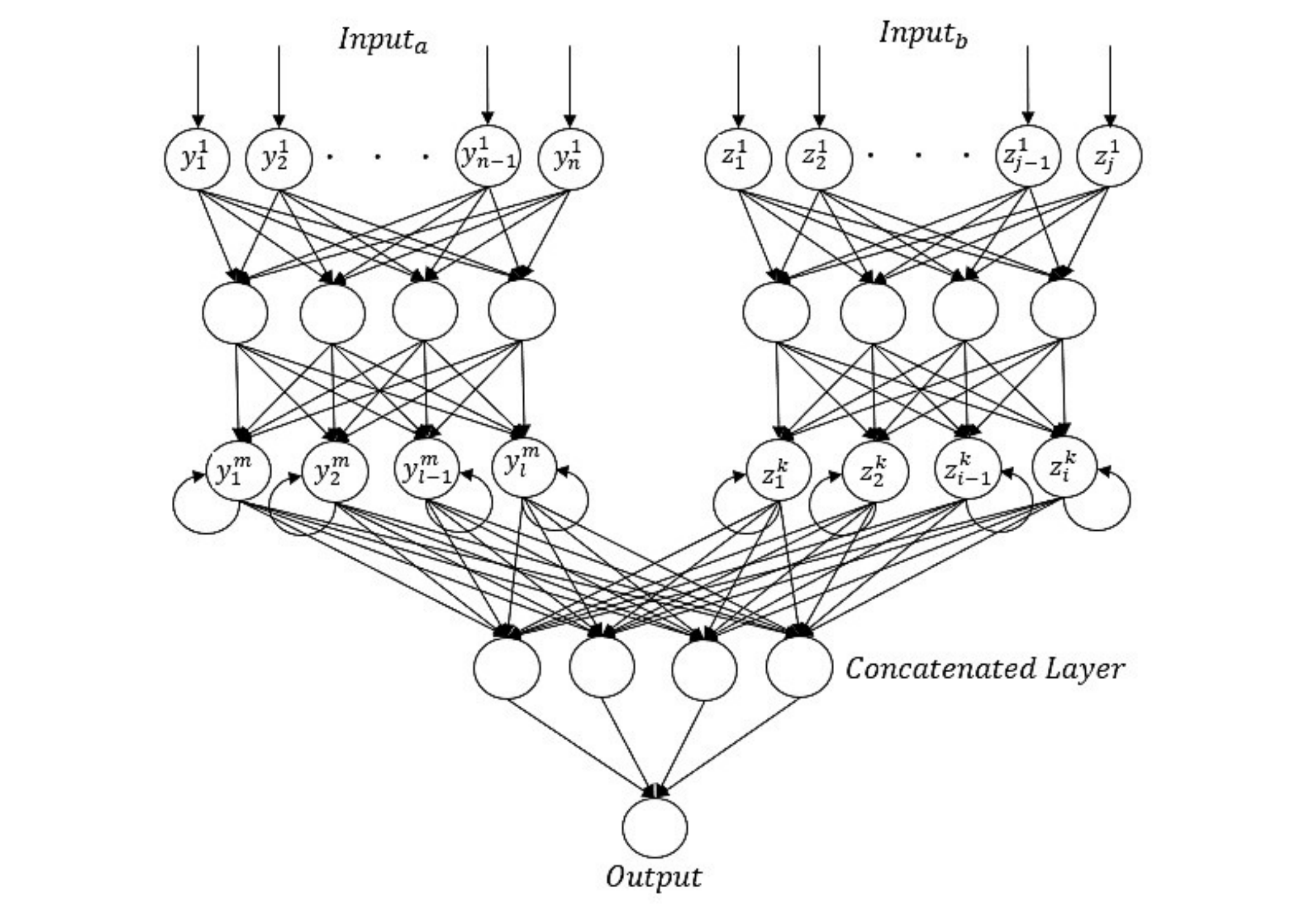}
    \caption{Diagram of a functional recurrent neural network, which consists of an input layer, a hidden layer and an output layer for two networks, a concatenated layer that takes the output from the last layer of the two parallel networks, and an output layer.}
    \label{fig: fun_rnn_diagram}
\end{figure}

\section{Estimation Process and Results}
\label{section4}
To evaluate the performance of the system after the training process we choose a group of O-type stars spectra observed at the National Astronomical Observatory of San Pedro Martir over different observing seasons. These spectra were taken using the 2.1M telescope with an Echelle spectograph. On Table \ref{tab: Obs Specs} the observed spectra and the SNR for every stellar spectra are shown.

{\renewcommand{\arraystretch}{1.2}
\begin{table}
\setlength\tabcolsep{4pt}
    \centering
    \begin{tabular}{c c c}
    Star & Type & SNR  \\
    \hline
        HD216898 & O9V & 58.78 
        \\
        HD108 & O4-8f?p & 85.81
        \\
        HD203064 & O7.5IIIn & 189.30 
        \\
        HD216532 & O8.5V & 84.05 
        \\
        HD15558 & O4.5III & 103.01 
        \\
        HD15570 & O4If & 108.47 
        \\
        HD15629 & O4.5V & 89.40 
        \\
        HD190864 & O6.5III & 18.27 
        \\
        HD16691 & O4If & 84.05 
        \\
    \end{tabular}
    \caption{Observed spectra and signal to noise ratio (SNR). The spectral type was extracted from "The SIMBAD astronomical database", \citet{simbad}}
    \label{tab: Obs Specs}
\end{table}}

In Table \ref{tab: Stars_Refs} we show a group of estimations of the same stars shown on Table \ref{tab: Obs Specs} published on previous works, using different methods. Additionally, it is important to mention that, there are missing data in the observed spectra for the stars: HD15558, HD15570, HD15629, and HD190864, on the spectral signal related to 3,800 \AA \ and less.

\begin{table*}
\setlength\tabcolsep{4pt}
    \centering
    \begin{tabular}{c c c c c}
    Star & $T_{eff}(K)$ & $log$ $g$ & $log(L/L\textsubscript{\(\odot\)})$ & Refs. \\
    \hline
        HD216898 & 34000 $\pm$ 1000  & 4.00 $\pm$ 0.10 & 4.73 $\pm$ 0.25 & \citet{HD216532_HD216898} \\
        & & & &\\
        HD108  & \makecell{35000 $\pm$ 2000  \\ 37000 $\pm$ 2000} &\makecell{ 3.50 $\pm$ 0.10 \\ 3.75 $\pm$ 0.10} & \makecell{5.70 $\pm$ 0.10  \\ 5.40 $\pm$ 0.10} &\makecell{ \citet{HD108_1} \\  \citet{HD108_3} }\\
        & & & &\\
        HD203064 & \makecell{35000 $\pm$ 1500  \\ 34500} & \makecell{3.73 $\pm$ 0.15 \\3.50} & \makecell{5.10  +0.20/-0.30 \\5.50}  &\makecell{ \citet{HD203064} \\ \citet{HD203064_HD15629_HD190864} } \\
         & & & &\\
        HD216532 & 33000 $\pm$ 2000 & 3.70 $\pm$ 0.20 & 4.79  $\pm$ 0.25 & \citet{HD216532_HD216898} \\
        & & & &\\
        HD15558 & \makecell{41000 \\ 46500} & \makecell{3.80 \\ 3.71 $\pm$ 0.10} & \makecell{5.93 \\ 6.16} & \makecell{ \citet{HD15558_HD15570_HD16691} \\ \citet{HD15558_HD15570_HD15629} } \\
        & & & &\\
        HD15570 & \makecell{38000 $\pm$ 1000 \\ 50000} & \makecell{3.51 $\pm$ 0.10 \\ 3.51 $\pm$ 0.10} & \makecell{5.94 $\pm$ 0.10\\ 6.44} & \makecell{ \citet{HD15558_HD15570_HD16691}\\ \citet{HD15558_HD15570_HD15629} } \\
        & & & &\\
        HD15629 & \makecell{40000 \\ 48000} & \makecell{3.71 \\ 3.81 $\pm$ 0.10} &\makecell{5.60 \\ 5.89} & \makecell{ \citet{HD203064_HD15629_HD190864} \\ \citet{HD15558_HD15570_HD15629}} \\
        \\
        HD190864 & 37000 & 3.55  & 5.41 &  \citet{HD203064_HD15629_HD190864}\\
        \\
        HD16691 & 41000 $\pm$ 1000 & 3.66 $\pm$ 0.10  & 5.94 $\pm$ 0.10 &  \citet{HD15558_HD15570_HD16691}\\
    \end{tabular}
    \caption{Fundamental parameters estimations from other authors.}
    \label{tab: Stars_Refs}
\end{table*}

\begin{figure}
    \centering
    \includegraphics[scale=0.13]{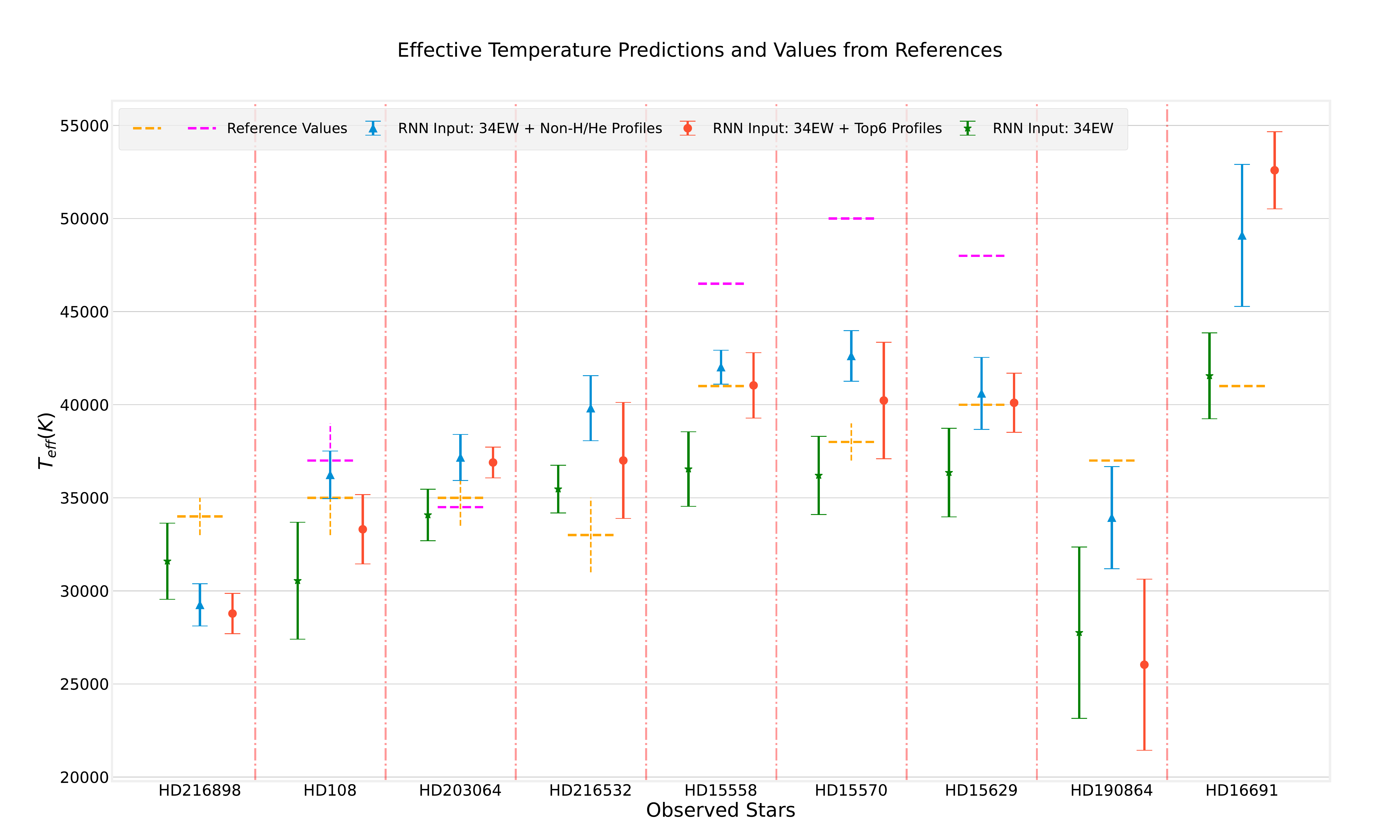}
    \caption{Effective Temperature predictions comparison over the 9 observed spectra using the 3 different RNNs approaches. Where the reference values without error bars doesn't have a uncertainty value on the published paper, and the error bars on the estimated values are the uncertainty of the RNN.}
    \label{fig: RNN_teff_pred}
\end{figure}

\begin{figure}
    \centering
    \includegraphics[scale=0.13]{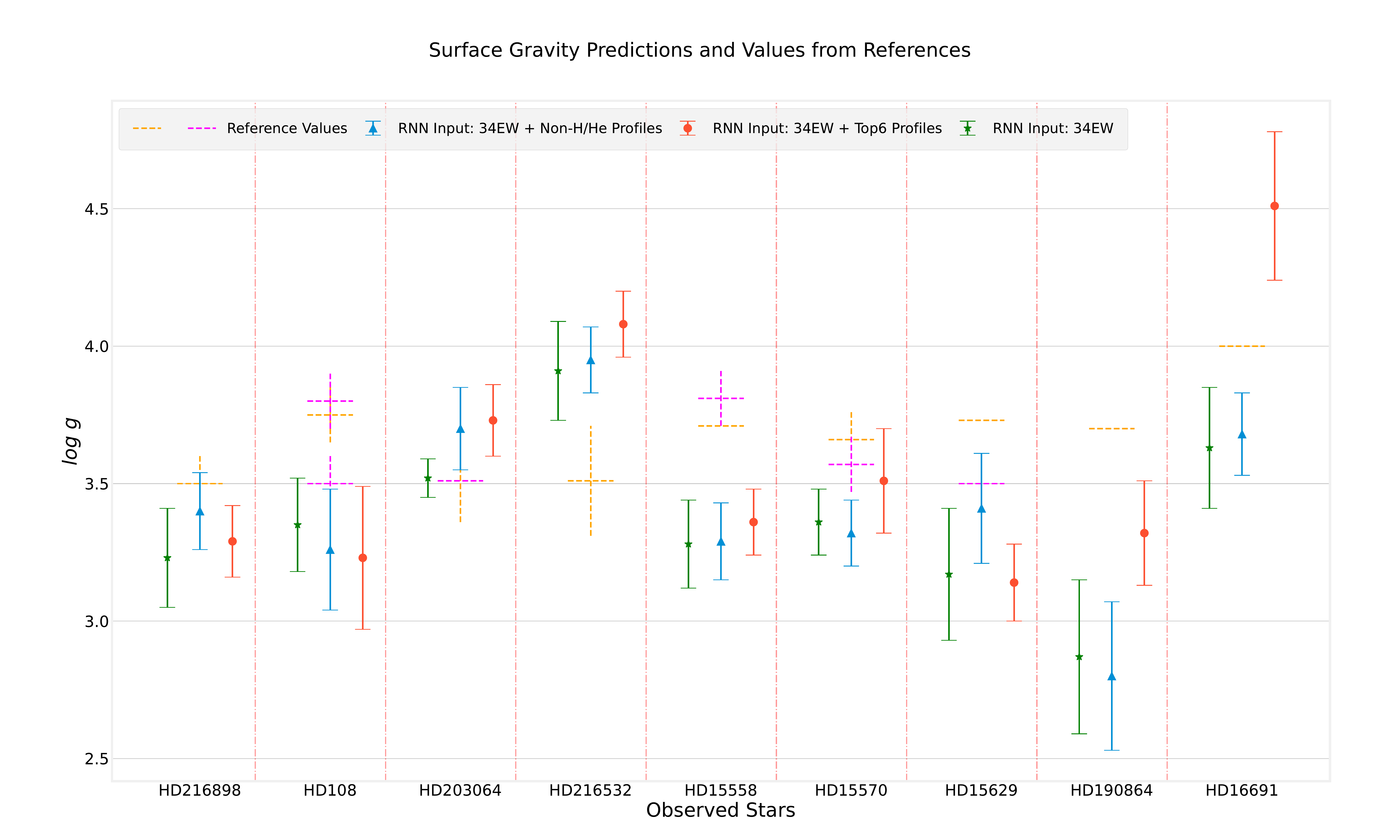}
    \caption{Surface Gravity predictions comparison over the 9 observed spectra using the 3 different RNNs approaches. Where the reference values without error bars doesn't have a uncertainty value on the published paper, and the error bars on the estimated values are the uncertainty of the RNN.}
    \label{fig: RNN_logg_pred}
\end{figure}

\begin{figure}
    \centering
    \includegraphics[scale=0.13]{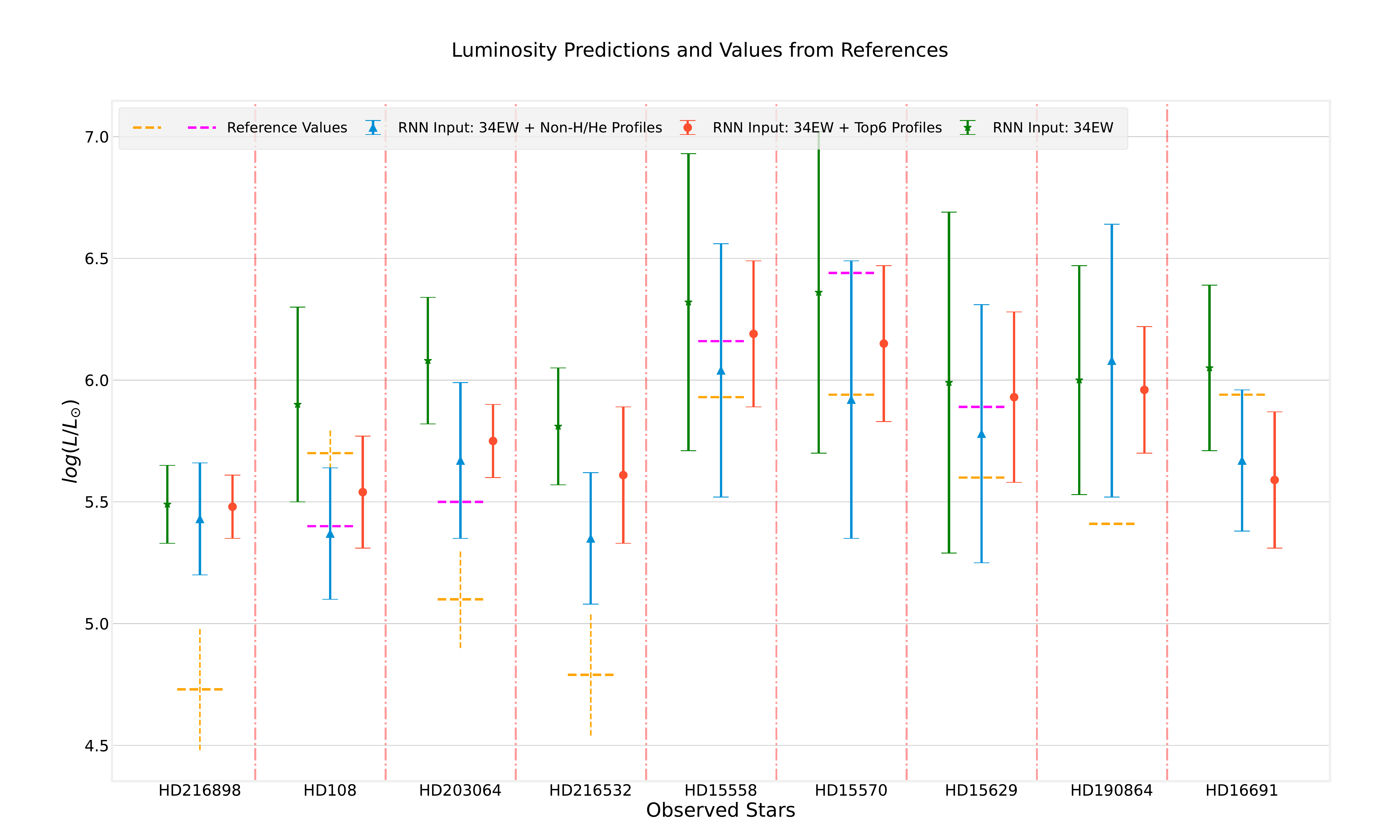}
    \caption{Luminosity predictions comparison over the 9 observed spectra using the 3 different RNNs approaches. Where the reference values without error bars doesn't have a uncertainty value on the published paper, and the error bars on the estimated values are the uncertainty of the RNN.}
    \label{fig: RNN_logl_pred}
\end{figure}

After neural network design, training, and validation, we create 10 different neural networks with the same structure put through the same training process in order to establish the uncertainty of every network due to the natural randomness of these deep learning algorithms. The results of this validation process are shown on Figures \ref{fig: RNN_teff_pred}, \ref{fig: RNN_logg_pred}, and \ref{fig: RNN_logl_pred}, where the uncertainties of the estimations are represented as error bars similar to the reference values from other authors.

In order to analyze the impact of the noise over the estimations of the different neural network structures, we compare the results with the estimations from other authors shown on Table \ref{tab: Stars_Refs}.

We show on Tables \ref{tab: gradmap_Teff}, \ref{tab: gradmap_logg}, \ref{tab: gradmap_logL}, 
the difference between the Effective Temperature, Surface Gravity, and Luminosity estimated by the RNN and the estimations shown on Table \ref{tab: Stars_Refs} made by previous authors. Every table shows the results using the distinct input tensors, The EW of the 34 lines and the top 6 profile lines with higher correlation, the EW of the 34 lines and the metals profile lines, and finally, the input tensor with only the EW of the 34 lines. Where the difference was calculated as relative to the reference values define in Eq. \ref{eq:diff}. 

These tables were ordered upward by S/N (from high to low noise) and using a gradient color where red is more noise, and most intense green, less noise. Also, the three columns of differences have a gradient color where red is more difference and most intense green less difference.

\begin{equation}\label{eq:diff}
    \Delta_{Relative} = \frac{|Reference_{val} - RNN_{val}|}{Reference_{val}},
\end{equation}

Where $Reference_{val}$ is the physical parameter from other authors and $RNN_{val}$ is the physical parameter estimated by the Recurrent Neural Network proposed in this work.

Based on the Eq. \ref{eq:diff}, the three comparative columns on the Tables \ref{tab: gradmap_Teff}, \ref{tab: gradmap_logg}, \ref{tab: gradmap_logL} are define in Table \ref{tab: diff_columns}.

{\renewcommand{\arraystretch}{1.8}
\begin{table}
\setlength\tabcolsep{15pt}
    \centering
    \begin{tabular}{c c}
     Column & Input Tensor\\
    \hline
       $\Delta_{Relative}^{(1)}$ & 34 EW \& Top 6 profile lines\\
       $\Delta_{Relative}^{(2)}$ & 34 EW \& Non-H/He profile lines \\
       $\Delta_{Relative}^{(3)}$ & 34 EW 
        \\
    \end{tabular}
    \caption{Comparative columns definition shown on Tables \ref{tab: gradmap_Teff}, \ref{tab: gradmap_logg}, \ref{tab: gradmap_logL}.}
    \label{tab: diff_columns}
\end{table}
}

\begin{table}
\setlength\tabcolsep{3.8pt}
        \begin{center}
            \begin{tabular}{ c c *{3}{|c} | c}
        Star & S/N	& \makecell{$\Delta_{Relative}^{(1)}$ \\ $T_{eff}(K)$ \vspace{0.1 cm} } & \makecell{$\Delta_{Relative}^{(2)}$ \\ $T_{eff}(K)$ \vspace{0.1 cm}} & \makecell{$\Delta_{Relative}^{(3)}$ \\ $T_{eff}(K)$ \vspace{0.1 cm}} & Reference	\\
        \hline
        HD190864 & 18.27 & \gradientteff{0.30} & \gradientteff{0.08} &	\gradientteff{0.25} & \citeyear{HD203064_HD15629_HD190864} \\
        HD216898 & 58.78 &	\gradientteff{0.15} & \gradientteff{0.14} & \gradientteff{0.07} & \citeyear{HD216532_HD216898}  \\
        HD216532 & 84.05 &	\gradientteff{0.12} & \gradientteff{0.21} &	\gradientteff{0.07} & \citeyear{HD216532_HD216898}\\
        HD16691 & 84.05 &	\gradientteff{0.28} & \gradientteff{0.20} &	\gradientteff{0.01} & \citeyear{HD15558_HD15570_HD16691}\\
        HD108 & 85.81 &	\gradientteff{0.05} & \gradientteff{0.04} &	\gradientteff{0.13} & 
        \citeyear{HD108_1}\\
        HD108 & 85.81 & \gradientteff{0.05} & \gradientteff{0.04} &	\gradientteff{0.13} & \citeyear{HD108_2}\\
        HD108 & 85.81 &	\gradientteff{0.10} & \gradientteff{0.02} &	\gradientteff{0.17} & \citeyear{HD108_3}\\
        HD15629 & 89.40 &	\gradientteff{0.00} & \gradientteff{0.02} &	\gradientteff{0.09} & \citeyear{HD203064_HD15629_HD190864}\\
        HD15629 & 89.40 &	\gradientteff{0.16} & \gradientteff{0.15} &	\gradientteff{0.24} & \citeyear{HD15558_HD15570_HD15629}  \\
        HD15558 & 103.01 & \gradientteff{0.00} & \gradientteff{0.02} & \gradientteff{0.11} &  \citeyear{HD15558_HD15570_HD16691} \\
        HD15558 & 103.01 & \gradientteff{0.12} & \gradientteff{0.10} & \gradientteff{0.21} &  \citeyear{HD15558_HD15570_HD15629}\\
        HD15570 & 108.47 & \gradientteff{0.06} & \gradientteff{0.12} & \gradientteff{0.05} & \citeyear{HD15558_HD15570_HD16691}\\
        HD15570 & 108.47 & \gradientteff{0.20} & \gradientteff{0.15} & \gradientteff{0.28} & \citeyear{HD15558_HD15570_HD15629}\\
        HD203064 & 189.30 & \gradientteff{0.05} & \gradientteff{0.06} & \gradientteff{0.03} & \citeyear{HD203064}\\
        HD203064 & 189.30 & \gradientteff{0.07} & \gradientteff{0.08} & \gradientteff{0.01} & \citeyear{HD203064_HD15629_HD190864}  \\
            \end{tabular}
    \end{center}
        \caption{Difference between the Surface Gravity estimated by the RNN and the reference from other authors (red is higher difference, green lower difference), ordered by S/N (in ascending order).}
    \label{tab: gradmap_Teff}
\end{table}

The Table \ref{tab: gradmap_Teff} shows that use the EW plus the Non-H/He profile lines the overall difference is lower compared to the other two input tensor versions. Additionally, non of the versions show a clear better estimation when the noise in the signal decrease, however the  $\Delta_{Relative}^{(2)}$ shows better results on the stellar spectrum with higher noise and the $\Delta_{Relative}^{(3)}$ shows better results on the 2nd (HD216898), 3rd (HD216532), and 4th (HD16691) star spectrum with more noise.

\begin{table}[!ht]
\setlength\tabcolsep{4pt}
        \begin{center}
            \begin{tabular}{ c c *{3}{|c} | c}
        Star & S/N	& \makecell{$\Delta_{Relative}^{(1)}$ \\ $log$ $g$ \vspace{0.1 cm} } & \makecell{$\Delta_{Relative}^{(2)}$ \\ $log$ $g$  \vspace{0.1 cm} } & \makecell{$\Delta_{Relative}^{(3)}$ \\ $log$ $g$  \vspace{0.1 cm}} & Reference	\\
        \hline
        HD190864 & 18.27 & \gradientlogg{0.11} & \gradientlogg{0.22} &	\gradientlogg{0.05} & \citeyear{HD203064_HD15629_HD190864} \\
        HD216898 & 58.78 &	\gradientlogg{0.15} & \gradientlogg{0.15} & \gradientlogg{0.20} & \citeyear{HD216532_HD216898}  \\
        HD216532 & 84.05 &	\gradientlogg{0.08} & \gradientlogg{0.07} &	\gradientlogg{0.10} & \citeyear{HD216532_HD216898}\\
        HD16691 & 84.05 &	\gradientlogg{0.04} & \gradientlogg{0.01} &	\gradientlogg{0.01} & \citeyear{HD15558_HD15570_HD16691}\\
        HD108 & 85.81 &	\gradientlogg{0.07} & \gradientlogg{0.07} &	\gradientlogg{0.08} & 
        \citeyear{HD108_1}\\
        HD108 & 85.81 & \gradientlogg{0.07} & \gradientlogg{0.07} &	\gradientlogg{0.08} & \citeyear{HD108_2}\\
        HD108 & 85.81 &	\gradientlogg{0.13} & \gradientlogg{0.13} &	\gradientlogg{0.14} & \citeyear{HD108_3}\\
        HD15629 & 89.40 &	\gradientlogg{0.11} & \gradientlogg{0.08} &	\gradientlogg{0.01} & \citeyear{HD203064_HD15629_HD190864}\\
        HD15629 & 89.40 &	\gradientlogg{0.13} & \gradientlogg{0.10} &	\gradientlogg{0.04} & \citeyear{HD15558_HD15570_HD15629}  \\
        HD15558 & 103.01 & \gradientlogg{0.13} & \gradientlogg{0.13} & \gradientlogg{0.10} &  \citeyear{HD15558_HD15570_HD16691} \\
        HD15558 & 103.01 & \gradientlogg{0.05} & \gradientlogg{0.06} & \gradientlogg{0.02} &  \citeyear{HD15558_HD15570_HD15629}\\
        HD15570 & 108.47 & \gradientlogg{0.03} & \gradientlogg{0.05} & \gradientlogg{0.11} & \citeyear{HD15558_HD15570_HD16691}\\
        HD15570 & 108.47 & \gradientlogg{0.03} & \gradientlogg{0.05} & \gradientlogg{0.11} & \citeyear{HD15558_HD15570_HD15629}\\
        HD203064 & 189.30 & \gradientlogg{0.02} & \gradientlogg{0.01} & \gradientlogg{0.08} & \citeyear{HD203064}\\
        HD203064 & 189.30 & \gradientlogg{0.04} & \gradientlogg{0.06} & \gradientlogg{0.02} & \citeyear{HD203064_HD15629_HD190864}  \\
            \end{tabular}
    \end{center}
        \caption{Difference between the Surface Gravity estimated by the RNN and the reference from other authors (red is higher difference, green lower difference), ordered by S/N (in ascending order).}
    \label{tab: gradmap_logg}
\end{table}

\begin{table}[!ht]
\setlength\tabcolsep{2.5pt}
        \begin{center}
            \begin{tabular}{ c c *{3}{|c} | c}
        Star & S/N	& \makecell{$\Delta_{Relative}^{(1)}$ \\ $log(L/L\textsubscript{\(\odot\)})$  \vspace{0.1 cm}} & \makecell{$\Delta_{Relative}^{(2)}$ \\ $log(L/L\textsubscript{\(\odot\)})$ \vspace{0.1 cm} } & \makecell{$\Delta_{Relative}^{(3)}$ \\ $log(L/L\textsubscript{\(\odot\)})$  \vspace{0.1 cm}} & Reference	\\
        \hline
        HD190864 & 18.27 & \gradientlogl{0.10} & \gradientlogl{0.12} &	\gradientlogl{0.11} & \citeyear{HD203064_HD15629_HD190864} \\
        HD216898 & 58.78 &	\gradientlogl{0.16} & \gradientlogl{0.15} & \gradientlogl{0.16} & \citeyear{HD216532_HD216898}  \\
        HD216532 & 84.05 &	\gradientlogl{0.17} & \gradientlogl{0.12} &	\gradientlogl{0.21} & \citeyear{HD216532_HD216898}\\
        HD16691 & 84.05 &	\gradientlogl{0.06} & \gradientlogl{0.05} &	\gradientlogl{0.02} & \citeyear{HD15558_HD15570_HD16691}\\
        HD108 & 85.81 &	\gradientlogl{0.03} & \gradientlogl{0.06} &	\gradientlogl{0.04} & 
        \citeyear{HD108_1}\\
        HD108 & 85.81 & \gradientlogl{0.03} & \gradientlogl{0.06} &	\gradientlogl{0.04} & \citeyear{HD108_2}\\
        HD108 & 85.81 &	\gradientlogl{0.03} & \gradientlogl{0.01} &	\gradientlogl{0.09} & \citeyear{HD108_3}\\
        HD15629 & 89.40 &	\gradientlogl{0.06} & \gradientlogl{0.03} &	\gradientlogl{0.07} & \citeyear{HD203064_HD15629_HD190864}\\
        HD15629 & 89.40 &	\gradientlogl{0.01} & \gradientlogl{0.02} &	\gradientlogl{0.02} & \citeyear{HD15558_HD15570_HD15629}  \\
        HD15558 & 103.01 & \gradientlogl{0.04} & \gradientlogl{0.02} & \gradientlogl{0.07} &  \citeyear{HD15558_HD15570_HD16691} \\
        HD15558 & 103.01 & \gradientlogl{0.00} & \gradientlogl{0.02} & \gradientlogl{0.03} &  \citeyear{HD15558_HD15570_HD15629}\\
        HD15570 & 108.47 & \gradientlogl{0.04} & \gradientlogl{0.00} & \gradientlogl{0.07} & \citeyear{HD15558_HD15570_HD16691}\\
        HD15570 & 108.47 & \gradientlogl{0.05} & \gradientlogl{0.08} & \gradientlogl{0.01} & \citeyear{HD15558_HD15570_HD15629}\\
        HD203064 & 189.30 & \gradientlogl{0.13} & \gradientlogl{0.11} & \gradientlogl{0.19} & \citeyear{HD203064}\\
        HD203064 & 189.30 & \gradientlogl{0.05} & \gradientlogl{0.03} & \gradientlogl{0.11} & \citeyear{HD203064_HD15629_HD190864}  \\
            \end{tabular}
    \end{center}
        \caption{Difference between the Luminosity estimated by the RNN and the reference from other authors (red is higher difference, green lower difference), ordered by S/N (in ascending order).}
    \label{tab: gradmap_logL}
\end{table}

After analyze the results on Figures \ref{fig: RNN_teff_pred}, \ref{fig: RNN_logg_pred}, and \ref{fig: RNN_logl_pred}, and compare with the difference on estimations with other authors shown on the gradient color Tables \ref{tab: gradmap_Teff}, \ref{tab: gradmap_logg}, \ref{tab: gradmap_logL}, it is clear that the improvement on the system by adding profile lines is not too high in all the observed spectra, while the difference of the computational resources to train the networks goes from processing  data tensors with a dimension of $34 \times N$, to a minimal of $34 \times 506 \times N$, and a max of $34 \times 904 \times N$, with $N$ as the total spectra samples. Moreover, the uncertainty for some observed spectra was close, as we can see on HD190864 and HD15629, where the 34 EW version of the system has very close estimations compared with the results on the 34 EW $+$ Non-H/He profiles lines on the same observed stellar spectra. Besides that, for the Effective Temperature estimations, the addition of profile lines got closer results to reference values over the 34 EW approach. Also, the uncertainty was lower in this case.
Additionally, the uncertainty on some observed spectra increases for all three approaches, over the observed spectra of HD15558, HD15570, HD15629, and HD190864 stars. A clear response to the missing data in the spectrum. In this field, the system with 34 EW $+$ top 6 profiles lines input tensor, got lower uncertainty on the estimations of Luminosity and Surface Gravity.

\section{Conclusions}
\label{section5}
We have developed a deep learning system using an RNN structure able to estimate stellar parameters with lower S/N observed spectra. In some cases, the uncertainty of the estimations are near to the published on previous works. 

The RNNs show good results with high and low levels of S/N. For the three input approaches the increase in the discrepancy respond to the decrease of the S/N level, however, we have some outliers in the estimations. Look in to with more detail, the stars mentioned before that does not have the spectrum complete show higher uncertainty compared with the other observations.

The different input data approaches show useful information about the possible requirements that the individual networks for every parameter could need. The RNN that uses the 34 EW \& the Non-H/He profile lines estimates closer values of $T_{eff}(K)$, $log(L/Lo)$, while for $log$ $g$, the RNN with the input tensor defines by the 34 EW \& Top 6 profile lines more correlated to this physical parameter, estimates closer values.

\subsection{\textit{Future Work}}
We think it is crucial to extend the observed spectra database with different levels of S/N and resolution in order to increase the analysis of the system capabilities, and improve the accuracy after enhance the training set and tuning the recurrent neural networks if its needed.
Design and develop a final system proposal in order to estimates more stellar parameters that the models already handle.
We are also open to defining different input tensors that can collect the most useful information of a full stellar spectrum and stills improve the system from a computational resources perspective.

\section*{Acknowledgments}
Based upon observations carried out at the Observatorio Astronómico Nacional on the Sierra San Pedro Mártir (OAN-SPM), Baja California, México.

\section*{Data Availability}
A set of models for O-type stars used in this article are available at \url{http://www.astroscu.unam.mx/atlas/}. Additionally, you can contact \href{celia.fierro.estrellas@gmail.com}{
Celia F. Fierro-Santillán} for the models availability. And
the trained model, training set, and test
data set will be shared on reasonable request to the author \href{miguel.flores.rod@gmail.com}{
Miguel Flores R}.



\bibliographystyle{mnras}
\bibliography{References} 

\begin{thebibliography}{}
\makeatletter
\relax
\def\mn@urlcharsother{\let\do\@makeother \do\$\do\&\do\#\do\^\do\_\do\%\do\~}
\def\mn@doi{\begingroup\mn@urlcharsother \@ifnextchar [ {\mn@doi@}
  {\mn@doi@[]}}
\def\mn@doi@[#1]#2{\def\@tempa{#1}\ifx\@tempa\@empty \href
  {http://dx.doi.org/#2} {doi:#2}\else \href {http://dx.doi.org/#2} {#1}\fi
  \endgroup}
\def\mn@eprint#1#2{\mn@eprint@#1:#2::\@nil}
\def\mn@eprint@arXiv#1{\href {http://arxiv.org/abs/#1} {{\tt arXiv:#1}}}
\def\mn@eprint@dblp#1{\href {http://dblp.uni-trier.de/rec/bibtex/#1.xml}
  {dblp:#1}}
\def\mn@eprint@#1:#2:#3:#4\@nil{\def\@tempa {#1}\def\@tempb {#2}\def\@tempc
  {#3}\ifx \@tempc \@empty \let \@tempc \@tempb \let \@tempb \@tempa \fi \ifx
  \@tempb \@empty \def\@tempb {arXiv}\fi \@ifundefined
  {mn@eprint@\@tempb}{\@tempb:\@tempc}{\expandafter \expandafter \csname
  mn@eprint@\@tempb\endcsname \expandafter{\@tempc}}}

\bibitem[\protect\citeauthoryear{{Bouret, J.-C.}, {Hillier, D. J.}, {Lanz, T.}
  \& {Fullerton, A. W.}}{{Bouret, J.-C.}
  et~al.}{2012}]{HD15558_HD15570_HD16691}
{Bouret, J.-C.} {Hillier, D. J.} {Lanz, T.}  {Fullerton, A. W.} 2012, \mn@doi
  [A\&A] {10.1051/0004-6361/201118594}, 544, A67

\bibitem[\protect\citeauthoryear{Bu, Kumar, Xie, Pan, Zhao  \& Wu}{Bu
  et~al.}{2020}]{Bu_2020}
Bu Y.,  Kumar Y.~B.,  Xie J.,  Pan J.,  Zhao G.,   Wu Y.,  2020, \mn@doi [The
  Astrophysical Journal Supplement Series] {10.3847/1538-4365/ab8bcd}, 249, 7

\bibitem[\protect\citeauthoryear{{Cazorla, Constantin}, {Naz\'e, Ya\"el},
  {Morel, Thierry}, {Georgy, Cyril}, {Godart, M\'elanie}  \& {Langer,
  Norbert}}{{Cazorla, Constantin} et~al.}{2017}]{HD203064}
{Cazorla, Constantin} {Naz\'e, Ya\"el} {Morel, Thierry} {Georgy, Cyril}
  {Godart, M\'elanie}  {Langer, Norbert} 2017, \mn@doi [A\&A]
  {10.1051/0004-6361/201730680}, 604, A123

\bibitem[\protect\citeauthoryear{{Dafonte, C.}, {Fustes, D.}, {Manteiga, M.},
  {Garabato, D.}, {\'Alvarez, M. A.}, {Ulla, A.}  \& {Allende Prieto,
  C.}}{{Dafonte, C.} et~al.}{2016}]{Dafonte_2016}
{Dafonte, C.} {Fustes, D.} {Manteiga, M.} {Garabato, D.} {\'Alvarez, M. A.}
  {Ulla, A.}  {Allende Prieto, C.} 2016, \mn@doi [A\&A]
  {10.1051/0004-6361/201527045}, 594, A68

\bibitem[\protect\citeauthoryear{Fierro-Santill{\'{a}}n, Zsarg{\'{o}}, Klapp,
  D{\'{\i}}az-Azuara, Arrieta, Arias  \& Sigalotti}{Fierro-Santill{\'{a}}n
  et~al.}{2018}]{Fierro_2018}
Fierro-Santill{\'{a}}n C.~R.,  Zsarg{\'{o}} J.,  Klapp J.,  D{\'{\i}}az-Azuara
  S.~A.,  Arrieta A.,  Arias L.,   Sigalotti L. D.~G.,  2018, \mn@doi [The
  Astrophysical Journal Supplement Series] {10.3847/1538-4365/aabd3a}, 236, 38

\bibitem[\protect\citeauthoryear{Herrero, Puls  \& Villamariz}{Herrero
  et~al.}{2000}]{HD15558_HD15570_HD15629}
Herrero A.,  Puls J.,   Villamariz M.,  2000, A\&A, 354, 193

\bibitem[\protect\citeauthoryear{{Jiang, Bin, et al.}}{{Jiang, Bin, et
  al.}}{2020}]{Bin_2020}
{Jiang, Bin, et al.} 2020, \mn@doi [Universe] {10.3390/universe6040060}, 6

\bibitem[\protect\citeauthoryear{{Koleva, M.}, {Prugniel, Ph.}, {Bouchard, A.}
  \& {Wu, Y.}}{{Koleva, M.} et~al.}{2009}]{ULySS}
{Koleva, M.} {Prugniel, Ph.} {Bouchard, A.}  {Wu, Y.} 2009, \mn@doi [A\&A]
  {10.1051/0004-6361/200811467}, 501, 1269

\bibitem[\protect\citeauthoryear{Li, Pan  \& Duan}{Li et~al.}{2017}]{Li_2017}
Li X.-R.,  Pan R.-Y.,   Duan F.-Q.,  2017, \mn@doi [Research in Astronomy and
  Astrophysics] {10.1088/1674-4527/17/4/36}, 17, 036

\bibitem[\protect\citeauthoryear{Li, Zeng, Wang, Du, Kong  \& Liao}{Li
  et~al.}{2022}]{2022_param_est}
Li X.,  Zeng S.,  Wang Z.,  Du B.,  Kong X.,   Liao C.,  2022, \mn@doi [Monthly
  Notices of the Royal Astronomical Society] {10.1093/mnras/stac1625}, 514,
  4588

\bibitem[\protect\citeauthoryear{{Marcolino, W. L. F.}, {Bouret, J.-C.},
  {Martins, F.}, {Hillier, D. J.}, {Lanz, T.}  \& {Escolano, C.}}{{Marcolino,
  W. L. F.} et~al.}{2009}]{HD216532_HD216898}
{Marcolino, W. L. F.} {Bouret, J.-C.} {Martins, F.} {Hillier, D. J.} {Lanz, T.}
   {Escolano, C.} 2009, \mn@doi [A\&A] {10.1051/0004-6361/200811289}, 498, 837

\bibitem[\protect\citeauthoryear{{Martins, F.}, {Escolano, C.}, {Wade, G. A.},
  {Donati, J. F.}, {Bouret, J. C.}  \& {MiMeS collaboration}}{{Martins, F.}
  et~al.}{2012}]{HD108_1}
{Martins, F.} {Escolano, C.} {Wade, G. A.} {Donati, J. F.} {Bouret, J. C.}
  {MiMeS collaboration} 2012, \mn@doi [A\&A] {10.1051/0004-6361/201118039},
  538, A29

\bibitem[\protect\citeauthoryear{Martins, Donati, Marcolino, Bouret, Wade,
  Escolano, Howarth  \& the MiMeS~Collaboration}{Martins
  et~al.}{2010}]{HD108_2}
Martins F.,  Donati J.-F.,  Marcolino W. L.~F.,  Bouret J.-C.,  Wade G.~A.,
  Escolano C.,  Howarth I.~D.,   the MiMeS~Collaboration 2010, \mn@doi [Monthly
  Notices of the Royal Astronomical Society]
  {10.1111/j.1365-2966.2010.17005.x}, 407, 1423

\bibitem[\protect\citeauthoryear{{Miguel Flores R.}, {Luis J. Corral}  \&
  {Celia R. Fierro-Santillán}}{{Miguel Flores R.} et~al.}{2021}]{FloresR}
{Miguel Flores R.} {Luis J. Corral}  {Celia R. Fierro-Santillán} 2021, \mn@doi
  [preprint (arXiv:2105.07110)] {10.48550/ARXIV.2105.07110}

\bibitem[\protect\citeauthoryear{Minglei, Jingchang, Zhenping, Xiaoming  \&
  Yude}{Minglei et~al.}{2020}]{Minglei_2017}
Minglei W.,  Jingchang P.,  Zhenping Y.,  Xiaoming K.,   Yude B.,  2020,
  \mn@doi [Optik] {10.1016/j.ijleo.2020.165004}, 218, 165004

\bibitem[\protect\citeauthoryear{{Navarro, S. G.}, {Corradi, R. L. M.}  \&
  {Mampaso, A.}}{{Navarro, S. G.} et~al.}{2012}]{Navarro_2012}
{Navarro, S. G.} {Corradi, R. L. M.}  {Mampaso, A.} 2012, \mn@doi [A\&A]
  {10.1051/0004-6361/201016422}, 538, A76

\bibitem[\protect\citeauthoryear{Nazé, Walborn  \& Martins}{Nazé
  et~al.}{2008}]{HD108_3}
Nazé Y.,  Walborn N.~R.,   Martins F.,  2008, {Revista mexicana de astronomía
  y astrofísica}, 44, 331

\bibitem[\protect\citeauthoryear{Recio-Blanco, Bijaoui  \&
  De~Laverny}{Recio-Blanco et~al.}{2006}]{Recio-Blanco_2006}
Recio-Blanco A.,  Bijaoui A.,   De~Laverny P.,  2006, \mn@doi [Monthly Notices
  of the Royal Astronomical Society] {10.1111/j.1365-2966.2006.10455.x}, 370,
  141

\bibitem[\protect\citeauthoryear{{Repolust, T.}, {Puls, J.}  \& {Herrero,
  A.}}{{Repolust, T.} et~al.}{2004}]{HD203064_HD15629_HD190864}
{Repolust, T.} {Puls, J.}  {Herrero, A.} 2004, \mn@doi [A\&A]
  {10.1051/0004-6361:20034594}, 415, 349

\bibitem[\protect\citeauthoryear{{Sander, A.}, {Shenar, T.}, {Hainich, R.},
  {G\'{\i}menez-Garc\'{\i}a, A.}, {Todt, H.}  \& {Hamann, W.-R.}}{{Sander, A.}
  et~al.}{2015}]{Sander_2015}
{Sander, A.} {Shenar, T.} {Hainich, R.} {G\'{\i}menez-Garc\'{\i}a, A.} {Todt,
  H.}  {Hamann, W.-R.} 2015, \mn@doi [A\&A] {10.1051/0004-6361/201425356}, 577,
  A13

\bibitem[\protect\citeauthoryear{Sharma, Kembhavi, Kembhavi, Sivarani, Abraham
  \& Vaghmare}{Sharma et~al.}{2019}]{Sharma_2019}
Sharma K.,  Kembhavi A.,  Kembhavi A.,  Sivarani T.,  Abraham S.,   Vaghmare
  K.,  2019, \mn@doi [Monthly Notices of the Royal Astronomical Society]
  {10.1093/mnras/stz3100}, 491, 2280

\bibitem[\protect\citeauthoryear{Sharma et~al.,}{Sharma
  et~al.}{2020}]{Sharma_2020}
Sharma K.,  et~al., 2020, \mn@doi [Monthly Notices of the Royal Astronomical
  Society] {10.1093/mnras/staa1809}, 496, 5002

\bibitem[\protect\citeauthoryear{Snider, Prieto, von Hippel, Beers, Sneden, Qu
  \& Rossi}{Snider et~al.}{2001}]{Snider_2001}
Snider S.,  Prieto C.~A.,  von Hippel T.,  Beers T.~C.,  Sneden C.,  Qu Y.,
  Rossi S.,  2001, \mn@doi [The Astrophysical Journal] {10.1086/323428}, 562,
  528

\bibitem[\protect\citeauthoryear{Teimoorinia}{Teimoorinia}{2012}]{Teimoorinia_2012}
Teimoorinia H.,  2012, \mn@doi [The Astronomical Journal]
  {10.1088/0004-6256/144/6/172}, 144, 172

\bibitem[\protect\citeauthoryear{Villavicencio-Arcadia, Navarro, Corral,
  Mart\'inez, Nigoche, Kemp  \& Ramos-Larios}{Villavicencio-Arcadia
  et~al.}{2020}]{Villavicencio_2020}
Villavicencio-Arcadia E.,  Navarro S.~G.,  Corral L.~J.,  Mart\'inez C.~A.,
  Nigoche A.,  Kemp S.~N.,   Ramos-Larios G.,  2020, \mn@doi [Mathematical
  Problems in Engineering] {10.1155/2020/1751932}, 2020, 15

\bibitem[\protect\citeauthoryear{{Wenger, M.} et~al.,}{{Wenger, M.}
  et~al.}{2000}]{simbad}
{Wenger, M.} et~al., 2000, \mn@doi [Astron. Astrophys. Suppl. Ser.]
  {10.1051/aas:2000332}, 143, 9

\bibitem[\protect\citeauthoryear{{Worley, C. C.}, {de Laverny, P.},
  {Recio-Blanco, A.}, {Hill, V.}, {Bijaoui, A.}  \& {Ordenovic, C.}}{{Worley,
  C. C.} et~al.}{2012}]{Worley_2012}
{Worley, C. C.} {de Laverny, P.} {Recio-Blanco, A.} {Hill, V.} {Bijaoui, A.}
  {Ordenovic, C.} 2012, \mn@doi [A\&A] {10.1051/0004-6361/201218829}, 542, A48

\bibitem[\protect\citeauthoryear{{Zsarg\'o, J., et al.}}{{Zsarg\'o, J., et
  al.}}{2020}]{Zsargo_2020}
{Zsarg\'o, J., et al.} 2020, \mn@doi [A\&A] {10.1051/0004-6361/202038066}, 643,
  A88

\makeatother
\end{thebibliography}







\bsp	
\label{lastpage}
\end{document}